\documentclass[aps,epsfig,prl,twocolumn,superscriptaddress,showpacs]{revtex4}
\usepackage{graphicx}
\usepackage{dcolumn}
\usepackage{bm}
\usepackage{lineno}
\usepackage{setspace}
\def\clevel{CL}               
\def\twb{\ensuremath{tWb}}               
\def\vma{\ensuremath{V\!-\!A}}               
\def\vpa{\ensuremath{V\!+\!A}}               
\def\wp{\ensuremath{W^+}}               
\def\fg{Fig.}
\def\fgs{Figs.}
\def\fp{\ensuremath{f^+}}               
\def\fl{\ensuremath{f^0}}               
\def\fm{\ensuremath{f^-}}               

\def\ppbar{\ensuremath{p\overline{p}}}             
\def\ttbar{\ensuremath{t\overline{t}}}             

\def\pt{$p_T$}                          
\def\met{\mbox{${\hbox{$E$\kern-0.6em\lower-.1ex\hbox{/}}}_T$}} 
\def\mex{\mbox{${\hbox{$E$\kern-0.6em\lower-.1ex\hbox{/}}}_x$}} 
\def\mey{\mbox{${\hbox{$E$\kern-0.6em\lower-.1ex\hbox{/}}}_y$}} 
\def\mexy{\mbox{${\hbox{$E$\kern-0.6em\lower-.1ex\hbox{/}}}_{x,y}$}} 

\def\invpb{${\rm pb^{-1}}$}
\def\gevcc{GeV/$c^2$}                   
\def\D0{D\O}                            
\Roman{table}
\begin{document}

\lefthyphenmin=2
\righthyphenmin=3
\topmargin -1cm



\title{
Search for \vpa\ current in top quark decay in \ppbar\ collisions at $\sqrt{s} = 1.96$ TeV 
}


\affiliation{Institute of Physics, Academia Sinica, Taipei, Taiwan 11529, Republic of China} 
\affiliation{Argonne National Laboratory, Argonne, Illinois 60439} 
\affiliation{Institut de Fisica d'Altes Energies, Universitat Autonoma de Barcelona, E-08193, Bellaterra (Barcelona), Spain} 
\affiliation{Baylor University, Waco, Texas  76798} 
\affiliation{Istituto Nazionale di Fisica Nucleare, University of Bologna, I-40127 Bologna, Italy} 
\affiliation{Brandeis University, Waltham, Massachusetts 02254} 
\affiliation{University of California, Davis, Davis, California  95616} 
\affiliation{University of California, Los Angeles, Los Angeles, California  90024} 
\affiliation{University of California, San Diego, La Jolla, California  92093} 
\affiliation{University of California, Santa Barbara, Santa Barbara, California 93106} 
\affiliation{Instituto de Fisica de Cantabria, CSIC-University of Cantabria, 39005 Santander, Spain} 
\affiliation{Carnegie Mellon University, Pittsburgh, PA  15213} 
\affiliation{Enrico Fermi Institute, University of Chicago, Chicago, Illinois 60637} 
\affiliation{Joint Institute for Nuclear Research, RU-141980 Dubna, Russia} 
\affiliation{Duke University, Durham, North Carolina  27708} 
\affiliation{Fermi National Accelerator Laboratory, Batavia, Illinois 60510} 
\affiliation{University of Florida, Gainesville, Florida  32611} 
\affiliation{Laboratori Nazionali di Frascati, Istituto Nazionale di Fisica Nucleare, I-00044 Frascati, Italy} 
\affiliation{University of Geneva, CH-1211 Geneva 4, Switzerland} 
\affiliation{Glasgow University, Glasgow G12 8QQ, United Kingdom} 
\affiliation{Harvard University, Cambridge, Massachusetts 02138} 
\affiliation{Division of High Energy Physics, Department of Physics, University of Helsinki and Helsinki Institute of Physics, FIN-00014, Helsinki, Finland} 
\affiliation{University of Illinois, Urbana, Illinois 61801} 
\affiliation{The Johns Hopkins University, Baltimore, Maryland 21218} 
\affiliation{Institut f\"{u}r Experimentelle Kernphysik, Universit\"{a}t Karlsruhe, 76128 Karlsruhe, Germany} 
\affiliation{High Energy Accelerator Research Organization (KEK), Tsukuba, Ibaraki 305, Japan} 
\affiliation{Center for High Energy Physics: Kyungpook National University, Taegu 702-701, Korea; Seoul National University, Seoul 151-742, Korea; and SungKyunKwan University, Suwon 440-746, Korea} 
\affiliation{Ernest Orlando Lawrence Berkeley National Laboratory, Berkeley, California 94720} 
\affiliation{University of Liverpool, Liverpool L69 7ZE, United Kingdom} 
\affiliation{University College London, London WC1E 6BT, United Kingdom} 
\affiliation{Centro de Investigaciones Energeticas Medioambientales y Tecnologicas, E-28040 Madrid, Spain} 
\affiliation{Massachusetts Institute of Technology, Cambridge, Massachusetts  02139} 
\affiliation{Institute of Particle Physics: McGill University, Montr\'{e}al, Canada H3A~2T8; and University of Toronto, Toronto, Canada M5S~1A7} 
\affiliation{University of Michigan, Ann Arbor, Michigan 48109} 
\affiliation{Michigan State University, East Lansing, Michigan  48824} 
\affiliation{Institution for Theoretical and Experimental Physics, ITEP, Moscow 117259, Russia} 
\affiliation{University of New Mexico, Albuquerque, New Mexico 87131} 
\affiliation{Northwestern University, Evanston, Illinois  60208} 
\affiliation{The Ohio State University, Columbus, Ohio  43210} 
\affiliation{Okayama University, Okayama 700-8530, Japan} 
\affiliation{Osaka City University, Osaka 588, Japan} 
\affiliation{University of Oxford, Oxford OX1 3RH, United Kingdom} 
\affiliation{University of Padova, Istituto Nazionale di Fisica Nucleare, Sezione di Padova-Trento, I-35131 Padova, Italy} 
\affiliation{LPNHE, Universite Pierre et Marie Curie/IN2P3-CNRS, UMR7585, Paris, F-75252 France} 
\affiliation{University of Pennsylvania, Philadelphia, Pennsylvania 19104} 
\affiliation{Istituto Nazionale di Fisica Nucleare Pisa, Universities of Pisa, Siena and Scuola Normale Superiore, I-56127 Pisa, Italy} 
\affiliation{University of Pittsburgh, Pittsburgh, Pennsylvania 15260} 
\affiliation{Purdue University, West Lafayette, Indiana 47907} 
\affiliation{University of Rochester, Rochester, New York 14627} 
\affiliation{The Rockefeller University, New York, New York 10021} 
\affiliation{Istituto Nazionale di Fisica Nucleare, Sezione di Roma 1, University of Rome ``La Sapienza," I-00185 Roma, Italy} 
\affiliation{Rutgers University, Piscataway, New Jersey 08855} 
\affiliation{Texas A\&M University, College Station, Texas 77843} 
\affiliation{Istituto Nazionale di Fisica Nucleare, University of Trieste/\ Udine, Italy} 
\affiliation{University of Tsukuba, Tsukuba, Ibaraki 305, Japan} 
\affiliation{Tufts University, Medford, Massachusetts 02155} 
\affiliation{Waseda University, Tokyo 169, Japan} 
\affiliation{Wayne State University, Detroit, Michigan  48201} 
\affiliation{University of Wisconsin, Madison, Wisconsin 53706} 
\affiliation{Yale University, New Haven, Connecticut 06520} 
\author{A.~Abulencia}
\affiliation{University of Illinois, Urbana, Illinois 61801}
\author{J.~Adelman}
\affiliation{Enrico Fermi Institute, University of Chicago, Chicago, Illinois 60637}
\author{T.~Affolder}
\affiliation{University of California, Santa Barbara, Santa Barbara, California 93106}
\author{T.~Akimoto}
\affiliation{University of Tsukuba, Tsukuba, Ibaraki 305, Japan}
\author{M.G.~Albrow}
\affiliation{Fermi National Accelerator Laboratory, Batavia, Illinois 60510}
\author{D.~Ambrose}
\affiliation{Fermi National Accelerator Laboratory, Batavia, Illinois 60510}
\author{S.~Amerio}
\affiliation{University of Padova, Istituto Nazionale di Fisica Nucleare, Sezione di Padova-Trento, I-35131 Padova, Italy}
\author{D.~Amidei}
\affiliation{University of Michigan, Ann Arbor, Michigan 48109}
\author{A.~Anastassov}
\affiliation{Rutgers University, Piscataway, New Jersey 08855}
\author{K.~Anikeev}
\affiliation{Fermi National Accelerator Laboratory, Batavia, Illinois 60510}
% ...
\collaboration{CDF Collaboration}
\noaffiliation



\begin{abstract}
We report an upper limit on the fraction of \vpa\ current, $f_{\vpa}$, in top quark decays, using
approximately 700~\invpb\ of \ppbar\ collisions  at $\sqrt{s}$=1.96 TeV acquired 
by the upgraded Collider Detector at Fermilab.
For the decay $t \to Wb \to \ell \nu b$ (where $\ell=e$ or $\mu$),
the invariant mass of the charged lepton and the bottom quark jet is sensitive
to the polarization of the $W$ boson.
We determine $f_{\vpa} = -0.06 \pm 0.25$ given a top quark mass of 175~\gevcc.
We set an upper limit on $f_{\vpa}$ of 0.29 at the 95\% confidence level,
which represents an improvement by a factor of two on the previous best direct limit.

\end{abstract}


\maketitle




The decay of the top quark, the most massive fundamental particle observed by experiment~\cite{cdfdiscovery,d0discovery}, is particularly interesting as a direct probe of the charged current weak 
interaction at the highest energy scale presently available.  
In the standard model, the spin-$\frac{1}{2}$\ top quark decays via 
the charged current weak interaction to a spin-$1$ \wp\ boson
and a spin-$\frac{1}{2}$\ $b$ quark~\cite{ccnote},
with a branching fraction above 99\% and width $\Gamma_{t}=1.4$~GeV~\cite{topwidth}
for a top mass of 175~GeV/$c^{2}$.   
The lifetime of the top quark, $\hbar/\Gamma_t \sim 5\times 10^{-25}$~s,
is an order of magnitude shorter than the typical strong interaction time-scale 
for binding of quarks into hadrons, $\hbar/\Lambda_{\mathrm QCD} \sim 3 \times 10^{-24}$~s.
Therefore, the top quark decays before hadronization, and the spin information is directly transferred to the decay products. 
In the limit $m_b \to 0$, the pure \vma\ theory of the weak interaction predicts 
that the $b$ quark has left-handed ($-1/2$) polarization (helicity) and 
the \wp\ boson can only have either longitudinal (zero) or left-handed ($-1$) polarization.
The right-handed ($+1$) polarization is forbidden.
The fraction \fl\ of \wp\ bosons with longitudinal polarization is predicted 
at leading order in perturbation theory to be $\fl = m^2_t/(2m^2_W+m^2_t)=0.70$~\cite{kane}.
The non-zero $b$ quark mass and the higher-order QCD and electroweak radiative corrections modify 
these predictions below the 1\% level~\cite{fischer,ewkcorr}.  
However, the presence of non-standard-model couplings in the 
\twb\ vertex could significantly modify the polarization of the top quark decay 
products~\cite{kane,theory:vpa1,theory:vpa2,theory:vpa3}.  
Previous results have either been limited by the small statistics of the 
top quark samples~\cite{kilminster, cdfrun2i,canelli,D0run2i} or have only set indirect limits~\cite{cleo}. 

In this Letter, we search for a \vpa\ current in top quark decay, while assuming that the \ttbar\ production mechanism
is in agreement with the standard model prediction.  We further assume the absence of couplings from magnetic moment interactions in the \twb\ interaction,
so that \fl\ is unchanged from 0.70~\cite{kane}.  
Then, the \vpa\ fraction $f_{\vpa}$ is related to the fraction \fp\ of right-handed \wp\ bosons by $f_{\vpa} = \fp / (1 - \fl)$, and
the \vma\ fraction $f_{\vma} \equiv 1-f_{\vpa}$ is related to the fraction \fm\ of left-handed \wp\ bosons by $f_{\vma} = \fm / (1 - \fl)$.
The \wp\ boson polarization can be inferred from the angular distribution of the charged lepton~\cite{leptonnote} 
in the decay $W^+ \to \ell^+ \nu$, 
\begin{linenomath*}
\[ \label{eq:dGamma_dCosTheta}
\frac{1}{\Gamma}\frac{d\Gamma}{d\cos\theta}
\hspace*{-0.65mm}=\hspace*{-0.35mm}\frac{3}{4}\hspace*{-0.35mm}(1\hspace*{-0.95mm}-\hspace*{-0.65mm}\cos^2\hspace*{-0.35mm}\theta)\hspace*{-0.35mm}\fl
\hspace*{-0.65mm}+\hspace*{-0.35mm}\frac{3}{8}(1\hspace*{-0.95mm}-\hspace*{-0.65mm}\cos\hspace*{-0.35mm}\theta)^{2}\hspace*{-0.35mm}\fm
\hspace*{-0.65mm}+\hspace*{-0.35mm}\frac{3}{8}(1\hspace*{-0.95mm}+\hspace*{-0.65mm}\cos\hspace*{-0.35mm}\theta)^{2}\hspace*{-0.35mm}\fp,
\]
\end{linenomath*}
where the angle $\theta$ is the polar angle of the charged lepton in the rest frame of the \wp\ boson. 
The $z$-axis is defined to be the direction of motion of the \wp\ boson in the rest frame of the top quark.  
We use the observable 
$M^2_{{\ell}b}$, the square of the invariant mass of the charged lepton and the jet from the $b$ quark,
which is related to $\cos\theta$\ by
\begin{linenomath*}
\[
M_{\ell b}^2 \simeq \frac{1}{2}(m_t^2-m_W^2)(1+\cos\theta).
\]
\end{linenomath*}
The relation is exact in the limit $m_b{\rightarrow}0$. 


This search is based on a data set with an integrated luminosity of approximately 700~\invpb\ 
acquired by the Collider Detector at Fermilab (CDF II)\cite{CDFII1} from \ppbar\ collisions at $\sqrt{s}=1.96$~TeV. 
A 96-layer outer drift chamber~\cite{COT} reconstructs the trajectories of charged particles 
and measures their momenta in the region $|\eta|<1$~\cite{geomnote}. 
An eight-layer silicon microstrip detector~\cite{SVXNIM} provides precision
tracking information in the region $|\eta|<2$ to identify displaced vertices associated with $b$ hadron decays.
The entire tracking volume is located inside a 1.4~T magnetic field. 
Electromagnetic and hadronic calorimeters measure the energies
of particle showers.  Drift chambers and scintillation counters 
provide muon identification outside the calorimeters.
Gas Cherenkov counters~\cite{clc} determine the luminosity.   
The data are collected with an inclusive lepton trigger that
requires an electron (muon) with $|\eta|<1$ and 
$E_{T}>18$~GeV ($P_{T}>18$~GeV/$c$)~\cite{geomnote}. 


We study three independent data samples enriched in \ttbar\ events.
Two of the data samples are in the lepton$+$jets channel, with 
\ttbar$\rightarrow W^{+}bW^{-}\bar{b}$ events where one of the $W$
bosons decays hadronically and the other leptonically.  
The lepton$+$jets event selection requires one isolated lepton with $E_T>20$~GeV, 
$\met>20$~GeV~\cite{geomnote}, 
at least three jets with $E_T>15$~GeV, and one or two $b$-tagged jets. 
More details on the selection, the $b$-tagging procedure, and the 
sample composition can be found in Ref.~\cite{ttbarxslj}.
We model the hard \ttbar\ process with the Monte Carlo (MC) event generator {\sc alpgen}~\cite{alpgen} with {\sc cteq5l}~\cite{cteq} parton densities and {\sc pythia} \cite{pythia} for hadronization, under the assumption that the top quark mass is 175~\gevcc.
We simulate the detector response using {\sc geant}~\cite{geant1,geant2}.
For \ttbar\ production with \vma\ top quark decay, we estimate a selection efficiency, including the branching fraction, 
of $\mathcal{A}_{\vma}=3.4\% (1.2\%)$ for events with one (two) $b$-tagged jets.
Due to the lower average \pt\ of the charged lepton for \vma,
this is a factor 0.92 below the efficiency for \vpa.

For the lepton$+$jets sample with a single $b$-tagged jet,
the $b$-tagged jet is from the same top quark decay as the charged lepton in approximately
half of the \ttbar\ events.  
The background $M_{\ell b}^2$ distribution is a combination of 85$\%$ $W$+jets, modeled by {\sc alpgen}
$Wb\bar{b}$, and 15$\%$ multi-jet events, modeled by non-isolated lepton$+$jets data events.
Background-dominated data samples with only one jet or only two jets
are consistent, in terms of both the rate and the shape of the $M^{2}_{lb}$\ distribution, with our model of the background. 
In 695 \invpb, we observe 304 candidates with a total expected background of 88$\pm$11 events. 


For the lepton$+$jets sample with two $b$-tagged jets, 
the two possible $M_{lb}^{2}$\ values of the charged lepton with either
the highest or the second highest $E_{T}$\ $b$-tagged jet are used to construct 
a 2-D distribution.
In this way, we keep both the correct and incorrect combinations, and account for their correlation.
The background is modeled by {\sc alpgen} $Wb\bar{b}$; here the multi-jet background is negligible.
Non-uniform binning was applied in the 2-D $M_{lb}^2$ distributions 
in order to ensure sufficient MC events in each bin.
In 695 \invpb, we find 75 candidates with a total expected background of 9$\pm$2 events. 

The third sample is in the dilepton channel, with \ttbar$\rightarrow W^{+}bW^{-}\bar{b}$ events where both $W$ bosons decay leptonically.
The dilepton event selection requires two
identified leptons with opposite electric charge and $E_T>20$~GeV, $\met>25$~GeV, and at least two
jets with $E_T>15$~GeV\@. More details on the selection and the sample composition can be found
in Ref.~\cite{ttbarxsdil}.  
For \ttbar\ production with \vma\ top quark decay, 
modeled by {\sc alpgen} as described above, we estimate a selection efficiency, including the branching fraction, of $\mathcal{A}_{\vma}=0.72$\%,
a factor 0.88 below the efficiency for \vpa.
The two possible $M_{lb}^{2}$\ values for a charged lepton with either
the highest or the second highest $E_{T}$\ jet, 
assumed to be produced by the fragmentation of the $b$ quarks, are used to construct 
a 2-D distribution.  As we can reconstruct $M_{lb}^2$\ from
the top quark decay and from the anti-top quark decay, we make one entry for each charged lepton.
The effect of the correlation between the 
spins of the top quark and the anti-top quark is negligible here.
Again, non-uniform binning in the 2-D $M_{lb}^2$ distributions is applied.
The background $M_{\ell b}^2$ distribution is the
combination of three background types: approximately 50$\%$ from $Z/{\gamma}^{*} \to \ell^+ \ell^-$\ 
with associated jets, 30$\%$ from $W \to \ell\nu$ with associated jets where a jet is misidentified as a lepton, and 20$\%$ from massive diboson pairs, $WW$/$WZ$. The $Z/{\gamma}^{*}$\ and diboson background
$M_{lb}^{2}$\ distributions are modeled by {\sc alpgen}\@. The misidentified lepton background is 
based on inclusive lepton trigger data where the 
second lepton is instead a jet (charged particle track) weighted by a probability for misidentification as an electron (muon).
A background-dominated data sample with only one jet 
is consistent, in terms of both the rate and the shape of the $M^{2}_{lb}$\ distribution, with our model of the background. 
In 750 \invpb, we observe 64 candidates (12 $ee$, 24 ${\mu}{\mu}$, and 28 $e\mu$) with a total estimated background of 20$\pm$4 events.

The fraction $f_{\vpa}$ is estimated by comparing the $M_{\ell b}^2$ distribution in data with 
parent $M_{\ell b}^2$ distributions for \ttbar\ production with \vma\ top quark decay ($f_{\vpa}=0.0$), 
\ttbar\ production with \vpa\ top quark decay ($f_{\vpa}=1.0$), and backgrounds.
A binned log likelihood fit procedure is used to extract the parameter of interest, $f_{\vpa}$.  
We represent the imperfectly known accepted background cross section for each sample, $\sigma_{bg}$, and the \ttbar\ cross
section~\cite{cacciari,kidonakis}, $\sigma_{t\bar{t}}$, by nuisance parameters.
The analytic expression for the likelihood for each sample,  
\begin{linenomath*}
\begin{eqnarray}\label{eq:poisson}
\mathcal{L} \hspace*{-0.65mm} &=& \hspace*{-0.65mm} \Big[\prod^{N}_{i=0} P(n_i, \mu_i)\Big]\hspace*{-0.65mm} 
\times \hspace*{-0.65mm} G(\sigma_{bg},\delta_{\sigma_{bg}}) \hspace*{-0.65mm}
\times \hspace*{-0.65mm} G(\sigma_{t\bar{t}},\delta_{\sigma_{t\bar{t}}}),
\label{eq:likesample}
\end{eqnarray}
\end{linenomath*}
is the product over all $N$ bins in $M_{\ell b}^2$ of the Poisson probabilities
of observing $n_i$ entries in a given bin $i$, where the average expected bin content is ${\mu}_i$, 
and the Gaussian constraints on the estimated background and the predicted \ttbar\ production cross sections,
as shown in Table~\ref{tab:nuis}.  The ${\mu}_i$ are given by:
\begin{linenomath*}
\begin{eqnarray}
\label{eq:mui}
\mu_i=N^{data}\Big[x_{\vpa}\hat{T}^i_{\vpa}+x_{\vma}\hat{T}^i_{\vma}+x_{bg}\hat{T}^i_{bg} \Big], \\
\label{eq:xfrac0}
x_{\vpa}=\frac{f_{\vpa}\mathcal{A}_{\vpa}\sigma_{t\bar{t}}}{\sigma_{bg}+\sigma_{t\bar{t}}[\mathcal{A}_{\vpa}f_{\vpa}+\mathcal{A}_{\vma}(1-f_{\vpa})]},\\
\label{eq:xfrac1}
x_{\vma}=\frac{(1-f_{\vpa})\mathcal{A}_{\vma}\sigma_{t\bar{t}}}{\sigma_{bg}+\sigma_{t\bar{t}}[\mathcal{A}_{\vpa} f_{\vpa}+\mathcal{A}_{\vma}(1-f_{\vpa})]},\\
x_{bg}=\frac{\sigma_{bg}}{\sigma_{bg}+\sigma_{t\bar{t}}[\mathcal{A}_{\vpa} f_{\vpa}+\mathcal{A}_{\vma}(1-f_{\vpa})]}.
\label{eq:xfrac2}
\end{eqnarray}
\end{linenomath*}
Here, $N^{data}$ is the total number of observed events for the sample. The
$x_{\vpa},x_{\vma}$, and $x_{bg}$ are the fractions of
\ttbar\ production with \vpa\ top quark decay, \ttbar\ production with \vma\ top quark decay, and background, respectively. The
$\hat{T}^i_{\vpa}$, $\hat{T}^i_{\vma}$, and $\hat{T}^i_{bg}$ are the
probabilities for an event to occupy bin $i$ of the corresponding $M_{\ell b}^2$
distribution.
Note that $\sum_{i} \hat{T}^{i}=1.0$.  The combined likelihood is the product of the likelihoods of
the three samples, but with one common Gaussian constraint on the \ttbar\ cross section.

\begin{table}[bt]
\caption{The input values for the nuisance parameters, and the values from the best fit to the combined samples.  \label{tab:nuis}}
\begin{center}
\begin{ruledtabular}
\begin{tabular}{|l|c|c|}
Nuisance parameter    & Input (pb) & Fit (pb)\\
\hline
${\sigma}_{t\bar{t}}$                       & 6.7 $\pm$ 1.0 & 7.3 $\pm$ 0.9\\ 
$\sigma_\mathrm{bg}$ lepton+jets 1 $b$-tag  & 0.156 $\pm$ 0.017 & 0.154 $\pm$ 0.016\\
$\sigma_\mathrm{bg}$ lepton+jets 2 $b$-tag  & 0.013 $\pm$ 0.002 & 0.013 $\pm$ 0.002\\ 
$\sigma_\mathrm{bg}$ dilepton  & 0.026 $\pm$ 0.006 & 0.022 $\pm$ 0.006\\ 
\end{tabular}
\end{ruledtabular}
\end{center}
\end{table}


The robustness of the fitting procedure has been tested with
pseudo-experiments. For a given pseudo-experiment and a particular sample, the number of observed data events, $N^{data}$, is distributed in three categories
(\ttbar\ production with \vpa\ top quark decay, \ttbar\ production with \vma\ top quark decay, and background) as multinomial
deviates according to their expected fractions. These fractions are first varied according to Gaussian distributions to incorporate the uncertainties in the 
background and the \ttbar\ cross sections.
The events are generated from the relevant $M_{\ell b}^2$ parent distribution
for each category. 
The hypotheses that $f_{\vpa}=0.0, 0.1,\ldots, 1.0$\ are studied for 2000 pseudo-experiments 
for all samples combined, as well for the three samples separately.
In all cases, the
fit is unbiased and stable.  An expected statistical uncertainty of 0.22 on $f_{\vpa}$ is found for the
combined case, while for the separate samples we found 
0.36, 0.41 and 0.49 for lepton$+$jets single $b$-tagged events, lepton$+$jets 
double $b$-tagged events, and dilepton events, respectively.

The estimates of the systematic uncertainties on the measured value for $f_{\vpa}$ are
shown in Table \ref{table:systf} for all samples combined. The leading sources of systematic uncertainty arise from uncertainties on the measured jet energy~\cite{jetenergy}, the background shape and normalization,
and limited MC statistics.  We determine all systematic uncertainties by performing
pseudo-experiments in which the systematic
parameter in question is varied and
the resulting simulated data are fit to the default parent distributions.
All shifts are evaluated at $f_{\vpa}=0$ since none of the systematic
sources are related to the structure of the \twb\ vertex. 


\begin{table}[bt]
\caption{The systematic uncertainties on the measurement of $f_{\vpa}$ for all samples combined.
\label{table:systf} }
\begin{center}
\begin{ruledtabular}
\begin{tabular}{|l|c|}
Source & Uncertainty   \\
\hline
Jet energy   &  0.10 \\
Background modeling & 0.04 \\ 
MC statistics&  0.04\\
Initial/Final state QCD radiation    &   0.02   \\
Multiple \ppbar\ interactions   & 0.02  \\
$b$-tag efficiency($E_T$) & 0.02\\
MC generator  &  0.01 \\
Parton densities & 0.01\\
\hline\hline
Total & 0.12 \\
\end{tabular}
\end{ruledtabular}
\end{center}
\end{table}

\begin{figure}[bt]
\includegraphics[width=0.48\textwidth]{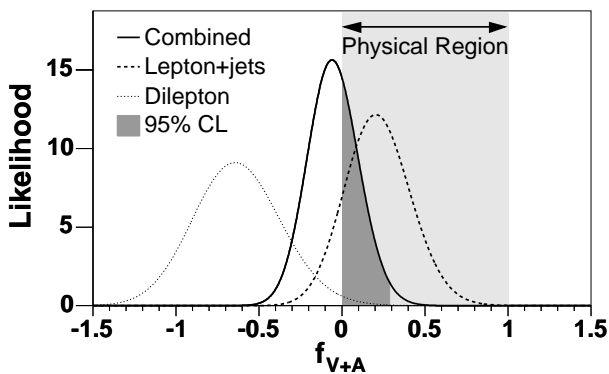}
\caption{Likelihood distribution (see Equation~\ref{eq:poisson}) for the lepton+jets and dilepton data samples separately and combined.}
\label{fig:llk_all}
\end{figure}
\begin{figure}[bt]
\includegraphics[width=0.48\textwidth]{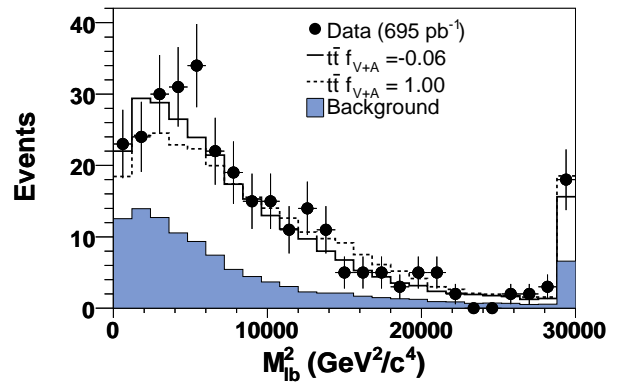}
\caption{The $M^{2}_{{\ell}b}$ distribution for the lepton$+$jets sample with a single $b$-tagged jet. }
\label{fig:mlbdatamcljets1b}
\end{figure}
\begin{figure}[bt]
\includegraphics[width=0.48\textwidth]{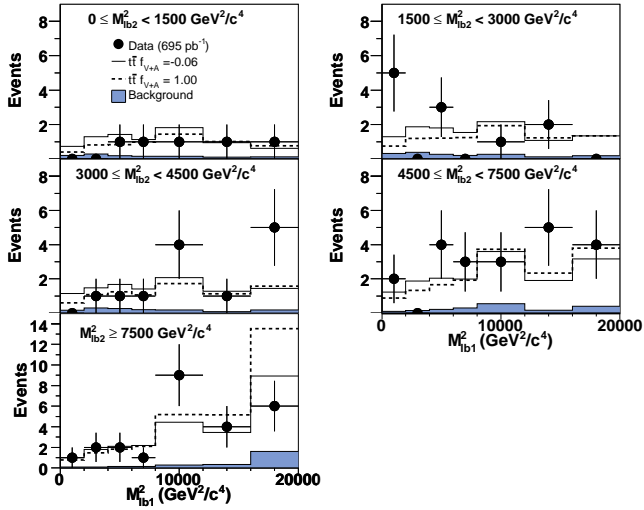}
\caption{The invariant mass squared of the charged lepton and the highest $E_T$ $b$-tagged jet, 
$M^{2}_{{\ell}b1}$, in five regions of the invariant mass squared of the charged lepton and 
the second highest $E_T$ $b$-tagged jet, $M^{2}_{{\ell}b2}$, for the lepton$+$jets sample with two $b$-tagged jets.}
\label{fig:mlbdatamcljets2b}
\end{figure}
\begin{figure}[bt]
\includegraphics[width=0.48\textwidth]{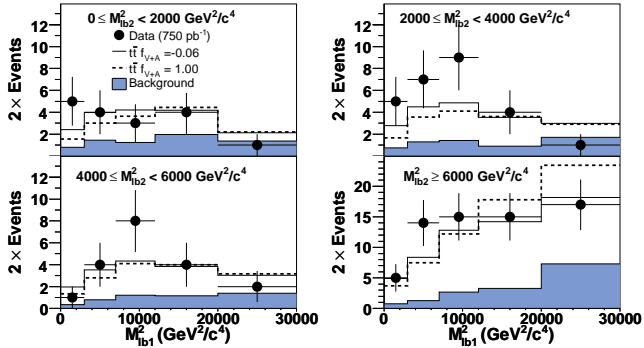}
\caption{The invariant mass squared of a charged lepton and the highest $E_T$ jet, 
$M^{2}_{{\ell}b1}$, in four regions of the invariant mass squared of the charged lepton and 
the second highest $E_T$ jet, $M^{2}_{{\ell}b2}$, for the dilepton sample.}
\label{fig:mlbdatamcdil}
\end{figure}

The maximum likelihood fit for the lepton$+$jets sample (single and double $b$-tagged) 
yields a value of $f_{\vpa}=0.21\pm 0.28$,
including the effects of the uncertainty on the background and \ttbar\ cross sections.
For the dilepton sample, we obtain $f_{\vpa}=-0.64\pm 0.37$.  
The probability to obtain a value smaller than the dilepton result is 10\% for the hypothesis $f_{\vpa}=0$. 
The lepton$+$jets and dilepton results are compatible at about 2.3 standard deviations. 
Combining all samples, and the total systematic uncertainty from  Table~\ref{table:systf}, 
the result for the fraction of \vpa\  current in top quark decay is
\begin{linenomath*}
\[
f_{\vpa}=-0.06\pm 0.25~(\textrm{stat.+syst.}).
\]
\end{linenomath*}
\noindent This value is in agreement with the standard model. Table~\ref{tab:nuis} summarizes the fitted values 
for the nuisance parameters. 
The likelihood distribution is shown in \fg~\ref{fig:llk_all}.
The good agreement in the $M^{2}_{{\ell}b}$ distribution between data and the best fit result for $f_{\vpa}$\ from combining all samples 
is shown in \fgs~\ref{fig:mlbdatamcljets1b}, \ref{fig:mlbdatamcljets2b}, and \ref{fig:mlbdatamcdil}, where the highest bins also contain overflow entries. For comparison, $f_{\vpa}=1.00$ is also shown.

In the absence of a signal, we evaluate an upper limit on $f_{\vpa}$\ using a Bayesian approach. The
profile likelihood function is first determined as a function of
$f_{\vpa}$, multiplied by a prior flat between 0.0 and 1.0, and normalized
to yield the posterior distribution for $f_{\vpa}$. 
The upper limit at 95$\%$ confidence level (\clevel) is formed by integrating the posterior from zero
to the value of $f_{\vpa}$ that yields 0.95 for the integral.
We verified that this approach yields proper frequentist coverage for 
$f_{\vpa} \leq 0.3$; a small correction, derived by the Neyman construction,
would be applied to any upper limit greater than 0.3 to restore coverage 
in the region $f_{\vpa} > 0.3$.
For the standard model, where $f_{\vpa}=0$, the median upper limit expected from combining all three samples is 0.38, 
with the 68\% interval from 0.2 to 0.6.
Combining all samples, we set an upper limit on the fraction of \vpa\ current in top quark decay of
\begin{linenomath*}
\[
f_{\vpa}<0.29 \textrm{ at 95\% \clevel}.
\]
\end{linenomath*}

\noindent This is an improvement by a factor of two on the previous best direct limit~\cite{kilminster}.
We estimate a $\pm 0.07$ ($\pm 0.09$) shift in the quoted upper limit (measurement) for $f_{\vpa}$ if 
the top quark mass is $\pm2.5$~\gevcc\ different from 175~\gevcc.
Converting the results for $f_{\vpa}$ to the fraction of right-handed $W^+$ bosons, 
we obtain $\fp = -0.02 \pm 0.07$\ and $\fp < 0.09$\ at 95\% \clevel.

We thank the Fermilab staff and the technical staffs of the participating institutions for their vital contributions. This work was supported by the U.S. Department of Energy and National Science Foundation; the Italian Istituto Nazionale di Fisica Nucleare; the Ministry of Education, Culture, Sports, Science and Technology of Japan; the Natural Sciences and Engineering Research Council of Canada; the National Science Council of the Republic of China; the Swiss National Science Foundation; the A.P. Sloan Foundation; the Bundesministerium f\"ur Bildung und Forschung, Germany; the Korean Science and Engineering Foundation and the Korean Research Foundation; the Particle Physics and Astronomy Research Council and the Royal Society, UK; the Institut National de Physique Nucleaire et Physique des Particules/CNRS; the Russian Foundation for Basic Research; the Comisi\'on Interministerial de Ciencia y Tecnolog\'{\i}a, Spain; the European Community's Human Potential Programme under contract HPRN-CT-2002-00292; and the Academy of Finland.

\bibliography{whel_prl}

\begin{thebibliography}{31}
\bibitem[{ccn()}]{ccnote}
\bibinfo{note}{{Charge conjugate decays, with replacement of left-handed by
  right-handed polarization, are implicit.}}
\bibitem[{lep()}]{leptonnote}
\bibinfo{note}{{We use lepton to denote an electron or a muon.}}
\bibitem[{geo()}]{geomnote}
\bibinfo{note}{{We use a cylindrical coordinate system where the $z$ axis is
  along the proton beam direction and $\theta$\ is the polar angle.
  Pseudorapidity is $\eta = -\ln\tan(\theta/2)$, while transverse momentum is
  $p_T = |p| \sin \theta$, and transverse energy is $E_T = E \sin \theta$.
  Missing transverse energy, \met, is defined as the magnitude of $-\sum_i
  E_T^i \hat{n}_i$, where $\hat{n}_i$\ is the unit vector in the azimuthal
  plane that points from the beam line to the $i^{\mathrm th}$ calorimeter
  tower.}}
\end{thebibliography}

\end{document}